\RequirePackage{luatex85}
\RequirePackage[OT1]{fontenc}
\documentclass[conference,letterpaper]{IEEEtran}
\usepackage[T1]{fontenc}

\usepackage[cmex10]{amsmath}
\usepackage{amssymb,amsfonts,amsthm}
\usepackage{array,multirow,graphicx}
\usepackage[dvipsnames]{xcolor}
\usepackage{tikz}
\usepackage{mathtools, nccmath}
\usetikzlibrary{positioning}
\usepackage{graphicx}
\usepackage{pgfplots} 
\usepackage{mathtools}
\pgfplotsset{compat=newest} 
\usepackage{textcomp}
\usepackage{cite,comment,enumitem}
\usepackage{bm}
\usepackage{accents}

\newcommand{\Prob}[1]{\textrm{Pr}\left[#1\right]}
\newcommand{\E}[1]{\textrm{E}\left[#1\right]}
\newcommand{\Eq}[1]{\textrm{E}_q\left[#1\right]}

\newcommand{\ddim}{{\rm{d}}}

\usepackage[top = 0.675 in, left = 0.65 in, right = 0.65 in, bottom = 0.85 in]{geometry}

\begin{document}

%%%%%%%%%%%%%%%%%%%%%%%%%%
%%%%%%%%%%%%%%%%%%%%%%%%%%
\title{Scalar Lattices and Probabilistic Shaping \\ for Dithered Wyner-Ziv Quantization}
%\thanks{Date of current version \today.}
%% Conference Veresion %%
\author{%
    \IEEEauthorblockN{M.~Yusuf~\c{S}ener\IEEEauthorrefmark{1}\IEEEauthorrefmark{2}, Gerhard~Kramer\IEEEauthorrefmark{1}, Shlomo Shamai (Shitz)\IEEEauthorrefmark{3}, and Wen Xu\IEEEauthorrefmark{2}}
    \IEEEauthorblockA{\IEEEauthorrefmark{1}%
    School of Computation, Information and Technology, Technical University of Munich, 80333 Munich, Germany}
   \IEEEauthorblockA{\IEEEauthorrefmark{2}%
    Munich Research Center, Huawei Technologies Duesseldorf GmbH, 80992 Munich, Germany}
   \IEEEauthorblockA{\IEEEauthorrefmark{3}%
    Dept. of Electrical and Computer Engineering, Technion—Israel Institute of Technology, Haifa 3200003, Israel}
    yusuf.sener@tum.de, gerhard.kramer@tum.de, sshlomo@ee.technion.ac.il, wen.xu@ieee.org
}
\begin{comment}
\thanks{Date of current version \today.}
%
\thanks{This work was supported by the German Research Foundation (DFG) via the German-Israeli Project Cooperation (DIP) under Project KR 3517/13-1 and SH 1937/1-1.}
%
\thanks{
M. Yusuf \c{S}ener is with the Institute for Communications Engineering, School of Computation, Information and Technology, Technical University of Munich, 80333 Munich, Germany, and also with the Munich Research Center, Huawei Technologies Duesseldorf GmbH, 80992 Munich, Germany (e-mail: yusuf.sener@tum.de).}
%
\thanks{Gerhard Kramer is with the Institute for Communications Engineering, School of Computation, Information and Technology, Technical University of Munich, 80333 Munich, Germany (e-mail: gerhard.kramer@tum.de).}
%
\thanks{Shlomo Shamai (Shitz) is with the Department of Electrical and Computer Engineering, Technion—Israel Institute of Technology, Haifa 3200003, Israel (e-mail:
sshlomo@ee.technion.ac.il).}
%
\thanks{Wen Xu is with the Munich Research Center, Huawei Technologies Duesseldorf GmbH, 80992 Munich, Germany (e-mail: wen.xu@ieee.org).}
\end{comment}

\maketitle

%%%%%%%%%%%%%%%%%%%%%%%%%
\begin{abstract}
Scalar lattice quantization with a modulo operator, dithering, and probabilistic shaping is applied to the Wyner-Ziv (WZ) problem with a Gaussian source and mean square error distortion. The method achieves the WZ rate-distortion pairs. The analysis is similar to that for dirty paper coding but requires additional steps to bound the distortion because the modulo shift is correlated with the source noise. The results extend to vector sources by reverse waterfilling on the spectrum of the covariance matrix of the source noise. Simulations with short polar codes illustrate the performance and compare with scalar quantizers and polar coded quantization without dithering.
\end{abstract}

%%%%%%%%%%%%%%%%%%%%%%%%%
\section{Introduction}
\label{sec:introduction}
Optimal lossy source coding with side information at the decoder combines Shannon coding~\cite{Shannon-48,Shannon-59} with Slepian-Wolf binning~\cite{Slepian-Wolf-IT73}. The best rate-distortion (RD) pairs are specified by the Wyner-Ziv (WZ) RD function~\cite{Wyner-Ziv-IT76,Wyner-IC78}; see~\cite[Ch.~15]{Cover06}. We study WZ coding for Gaussian vector sources and mean square error distortion; see~\cite{McDonald-Schultheiss-S64,Oohama-IT97,Puri-Ramchandran-All02,Oohma-IT05,Gastpar-Dragotti-Vetterli-IT06,Tian-Chen-IT09,Wang-Chen-IT14,Zahedi-ISIT14,gkagkos2024structural}.

WZ coding for real-valued sources can be implemented with multi-dimensional nested lattices~\cite{Zamir-Shamai-ISIT98,zamir2002nested,zamir14,Campello-IT19,Dongbo-IEEEA21}. Such lattices are helpful for many problems, e.g., dirty paper (DP) coding \cite{Costa-IT83} and Gaussian networks~\cite[Ch.~12]{zamir14}. However, the high-dimensional modulo operations of \cite{zamir2002nested} can be challenging to implement. Scalar shaping lattices are treated in~\cite[Ch.~6\ \&\ 9]{zamir14} but separated from quantization, and thus suboptimal.

A second approach applies nested linear codes directly to the DP and WZ problems. Polar codes~\cite{arikan2009channel} are well-suited because they enable practical joint coding and shaping~\cite{Korada-IT10,Sutter-ITW12,Honda-Yamamoto-IT13} and inherently permit multilevel coding \cite{Seidl-IT13}; cf.~\cite{Imai:IT77,wachsmann1999multilevel,Boecherer-WC17,Prinz-SPAWC17}. One can approximate optimal coding densities using amplitude shift keying (ASK) for DP coding or, similarly, an equally-spaced scalar quantizer for WZ coding~\cite{Eghbalian-Arani-ISWCS13,Liu-COMM16,Liu-COMM19,Liu-IT21,Liu-ISIT24,Jha-JSAIT22,Mondelli-IT18,Iscan-COMM18,Iscan-IA19,Iscan-TETT20,Wiegart-CL19,Boehnke-COMML20,Runge-ISIT22,Runge-ISIT24}. Multilevel binary polar codes are called polar coded modulation (PCM) in \cite{Seidl-IT13}, and polar lattices in \cite{Liu-COMM16,Liu-COMM19,Liu-IT21,Liu-ISIT24,Jha-JSAIT22} for ASK. For quantization, we call such methods polar coded quantization (PCQ).

A third approach \cite{Sener-CL21,Sener-CL24,Sener-ISIT24} uses scalar lattices, a modulo operator, and dithering. The primary purpose of this paper is to show this structure achieves the WZ curve. The proof is similar to \cite{Sener-CL24} but requires additional steps because the modulo shift is correlated with the source noise; see Sec.~\ref{subsec:distortion-scalar}. We also treat vector sources by reverse waterfilling on the spectrum of the covariance matrix of the source noise; see \cite[Fig.~10.7]{Cover06}. We remark dithering makes short block coding more challenging, see Sec.~\ref{subsec:scalar-simulations}, but has benefits for quantization and secrecy; see~\cite[Ch.~4]{zamir14},~\cite{Lipshitz-JAS92}. 

This paper is organized as follows. 
Sec.~\ref{sec:preliminaries} reviews notation, theory for Gaussian vectors, and several RD functions.
Sec.~\ref{sec:scalar-models} describes a coding structure with a modulo operator and dithering. Simulations illustrate the performance; the abbreviations ``PCQ'' and ``PCQmod'' refer to polar coding without and with the dithered modulo structure, respectively. 
Sec.~\ref{sec:vector-models} extends the results to vector sources.
Sec.~\ref{sec:conclusions} concludes the paper.

%%%%%%%%%%%%%%%%%%%%%%%%%
\section{Preliminaries}
\label{sec:preliminaries}

%%%%%%%%%%%%%%%%%%%%%%%%%
\subsection{Notation}
\label{subsec:notation}
Bold letters $\bm{x}=(x_1,\dots,x_{\ddim})^T$ refer to column vectors.
Let $\bm{1}_\ddim$ be the $\ddim$-dimensional all-ones vector and $I_\ddim$ be the $\ddim\times\ddim$ identity matrix.
%The determinant of the $d\times d$ matrix $Q$ is $|Q|$.
Let $\bm{x}\circ\bm{y}$ be the Hadamard (entry-by-entry) product of $\bm{x}$ and $\bm{y}$. Define the vector modulo operator
\begin{align}
    \bm{x}\ \textrm{mod}\ \bm{A} = \bm{x} - \bm{k} \circ\bm{A} 
\end{align}
where $\bm{A}$ has positive entries and $k_i$ is the unique integer for which $x_i-k_i A_i \in [-A_i/2,A_i/2)$ for $i=1,\dots,\ddim$.

Upper- and lowercase letters usually refer to random variables (RVs) and their realizations, e.g., $X$ is an RV, and $x$ is its realization. $P_X$ and $p_X$ are a probability mass function and density, respectively. We remove subscripts if the argument is the lowercase of the RV, e.g., $p(x)=p_X(x)$.
$\E{\bm{X}}$ and the random vector $\E{\bm{X}|\bm{Y}}$ are the expectations of $\bm{X}$ without and with conditioning on $\bm{Y}$, respectively. The corresponding covariance matrices are
\begin{align}
    Q_{\bm{X}} & = \E{(\bm{X}-\E{\bm{X}})(\bm{X}-\E{\bm{X}})^T} 
    \label{eq:cov-matrix} \\
    Q_{\bm{X}|\bm{Y}}
    & = \E{(\bm{X}-\E{\bm{X}|\bm{Y}})(\bm{X}-\E{\bm{X}|\bm{Y}})^T} .
    \label{eq:cond-cov-matrix}
\end{align}
For scalars, we write 
\eqref{eq:cov-matrix} and \eqref{eq:cond-cov-matrix} as $\sigma_x^2$ and  $\sigma_{x|y}^2$.
The notation $h(\bm{X})$, $I(\bm{X};\bm{Y})$, $I(\bm{X};\bm{Y}|\bm{Z})$ refers to the differential entropy of $\bm{X}$ and the mutual information of $\bm{X}$ and $\bm{Y}$ without and with conditioning on $\bm{Z}$, respectively. We write $\Eq{f(X)}:=\int_{\mathbb R} q(x) f(x)\, dx$ and $h_q(X):=\Eq{-\log q(X)}$.

%%%%%%%%%%%%%%%%%%%%%%%%%
\subsection{Gaussian Vectors}
\label{subsec:Gaussian-vectors}
We study jointly Gaussian $\bm{X},\bm{Y}$, perhaps with different dimensions, and with zero mean and joint covariance matrix $Q_{\bm{X},\bm{Y}}$. The conditional mean of $\bm{X}$ given $\bm{Y}$ is
\begin{align}
    \E{\bm{X}|\bm{Y}}
    & = \E{\bm{X}\bm{Y}^T} Q_{\bm{Y}}^{-1} \bm{Y}
    \label{eq:EXgY}
\end{align}
assuming $Q_{\bm{Y}}$ is invertible. One may write
\begin{align}
    \bm{X} = \E{\bm{X}|\bm{Y}} + \bm{Z}
    \label{eq:vec-Y-to-X-model}
\end{align}
where $\bm{Z}$ is independent of $\bm{Y}$ and
\begin{align}
    Q_{\bm{Z}} = Q_{\bm{X}|\bm{Y}}
    = Q_{\bm{X}} - \E{\bm{X}\bm{Y}^T} Q_{\bm{Y}}^{-1} \E{\bm{Y}\bm{X}^T} .
    \label{eq:QXgY}
\end{align}
Alternatively, we have
\begin{align}
    \bm{Y} = \E{\bm{Y}|\bm{X}} + {\bf \hat{Z}}
    \label{eq:vec-X-to-Y-model}
\end{align}
where ${\bf \hat{Z}}$ is independent of $\bm{X}$ and (see~\eqref{eq:QXgY})
\begin{align}
%    \E{\bm{Y}|\bm{X}}
%    & = \E{\bm{Y}\bm{X}^T} Q_{\bm{X}}^{-1} \bm{X}
%    \label{eq:EYgX} \\
    Q_{{\bf \hat{Z}}}
    & = Q_{\bm{Y}|\bm{X}}
    = Q_{\bm{Y}} - \E{\bm{Y}\bm{X}^T} Q_{\bm{X}}^{-1} \E{\bm{X}\bm{Y}^T}
    \label{eq:QYgX}
\end{align}
assuming $Q_{\bm{X}}$ is invertible.

%%%%%%%%%%%%%%%%%%%%%%%%%
\subsection{Rate-Distortion Theory}
\label{subsec:RD=theory}
Consider a source that outputs a string $X^n=X_1,\dots,X_n$ of independent and identically distributed (i.i.d.) symbols. The lossy source coding problem has a non-negative real-valued distortion function $d(.)$, an encoder mapping $X^n$ to $nR$ bits $W$, where $R$ is the rate, and a decoder mapping $W$ to a reconstruction $\hat X^n=\hat X_1,\dots,\hat X_n$. The goal is to design the encoder and decoder so the (random) empirical distortion
\begin{align}
    \Delta := \frac{1}{n} \sum\nolimits_{i=1}^n d(X_i,\hat{X}_i)
\end{align}
satisfies $\E{\Delta}\le\mathcal D$. We also study $\Prob{\Delta>\mathcal D}$, i.e., the probability of excess distortion for a specified $\mathcal D$.

The RD function is the infimum of $R$ as a function of $\mathcal D$. Shannon~\cite{Shannon-48,Shannon-59} showed that this function is
\begin{align}
    R(\mathcal D) = \min_{\E{d(X,\hat{X})}\le \mathcal D} I(X;\hat X).
\end{align}
For example, if $X$ is Gaussian and $d(x,\hat x) = (x-\hat x)^2$, then
\begin{align}
    R(\mathcal D) = \left\{ 
    \begin{array}{ll}
    \frac{1}{2}\log(\sigma_x^2\, \big/ \mathcal{D}), & 0< \mathcal{D} < \sigma_x^2 \\
    0, & \mathcal{D} \ge \sigma_x^2 .
    \end{array} \right.
\end{align}
For $0<\mathcal{D}<\sigma_x^2$, the optimal ``reverse'' channel is
\begin{align}
    X = \hat X + Z
    \label{eq:reverse-channel}
\end{align}
where $\hat X$ and $Z$ are independent and Gaussian with variances $\sigma_x^2-\mathcal{D}$ and $\mathcal{D}$, respectively. Applying \eqref{eq:vec-X-to-Y-model}-\eqref{eq:QYgX}, we have
\begin{align}
    \hat X = \left\{ 
    \begin{array}{ll}
    (1-\mathcal{D}/\sigma_x^2) X + \hat Z, & 0< \mathcal{D} < \sigma_x^2 \\
    0, & \mathcal{D} \ge \sigma_x^2
    \end{array} \right.
    \label{eq:forward-channel}
\end{align}
where $\hat Z$ is independent of $X$ and $\sigma_{\hat z}^2=(1-\mathcal{D}/\sigma_x^2) \mathcal{D}$.

%%%%%%%%%%%%%%%%%%%%%%%%%
\subsection{Conditional Rate-Distortion Theory}
\label{subsec:conditional-RD=theory}
Suppose the encoder and decoder have the side information $Y^n=Y_1,\dots,Y_n$ of i.i.d. symbols, i.e., the pairs $(X_i,Y_i)$ are i.i.d. for $i=1,\dots,n$. The conditional RD function is
\begin{align}
    R_{X|Y}(\mathcal D) = \min_{\E{d(X,\hat{X})}\le \mathcal D} I(X ; \hat X | Y).
    \label{eq:cond-RD-function}
\end{align}
For example, if $X,Y$ are jointly Gaussian and the distortion function is $d(x,\hat x)=(x-\hat x)^2$, then
\begin{align}
    R_{X|Y}(\mathcal D) = \left\{ 
    \begin{array}{ll}
    \frac{1}{2}\log(\sigma_{x|y}^2\, \big/ \mathcal D),
    & 0< \mathcal D < \sigma_{x|y}^2 \\
    0, & \mathcal D \ge \sigma_{x|y}^2 \;.
    \end{array} \right. 
\end{align}

More generally, suppose $\bm{X}$ and $\bm{Y}$ are jointly Gaussian and $d(\bm{x},\hat{\bm{x}})=\|\bm{x}-\hat{\bm{x}}\|^2$. Let $Q_{\bm{X}|\bm{Y}}=V \Lambda V^T$ be the eigenvalue decomposition of $Q_{\bm{X}|\bm{Y}}$, where $V$ is an orthogonal matrix and $\Lambda$ is non-negative with diagonal entries $\lambda_i$, $i=1,\dots,\ddim$, where $\ddim$ is the dimension of $\bm{X}$. Then, if $\mathcal D \ge \sum_{i=1}^\ddim \lambda_i$, we have $R_{\bm{X}|\bm{Y}}(\mathcal D)=0$. Otherwise, we have (see \cite[Ch.~10.3.3]{Cover06})
\begin{align}
    R_{\bm{X}|\bm{Y}}(\mathcal D) = 
    \sum\nolimits_{i: \lambda < \lambda_i} \frac{1}{2}\log(\lambda_i\, \big/ \lambda)
\end{align}
where the reverse-waterfilling level $\lambda>0$ is chosen so
\begin{align}
    \mathcal D = \sum\nolimits_{i=1}^\ddim \min(\lambda_i,\lambda).
\end{align}
The idea is that $i$-th encoder quantizes the $i$-th entry of 
\begin{align}
  \bm{X}_V=V^T (\bm{X} - \E{\bm{X}|\bm{Y}})
  \label{eq:XV}
\end{align}
of variance $\lambda_i$; observe that
\begin{align}
    Q_{\bm{X}_V} = V^T Q_{\bm{X}|\bm{Y}} V =\Lambda
    \label{eq:QXV}
\end{align}
so the entries of $\bm{X}_V$ are uncorrelated, and hence independent. If $\lambda_i$ is large, one transmits information about $X_{V,i}$; if $\lambda_i$ is small, one transmits no information. Reverse waterfilling states that the boundary between ``large'' and ``small'' is $\lambda$, and the best descriptions have distortion $\lambda$ if $\lambda<\lambda_i$. The reconstructions are as in \eqref{eq:forward-channel}, namely
\begin{align}
    \hat{X}_{V,i} = \left\{ 
    \begin{array}{ll}
    (1-\lambda/\lambda_i) X_{V,i} + \hat{Z}_i, & 0< \lambda < \lambda_i \\
    0, & \lambda \ge \lambda_i
    \end{array} \right.
    \label{eq:forward-channel-i}
\end{align}
where the $\hat{Z}_i$, $i=1,\dots,\ddim$, and $\bm{X}_V$ are mutually independent and
$\sigma_{\hat{z}_i}^2=(1-\lambda/\lambda_i) \lambda$. The output is $\hat{\bm{X}}=V \hat{\bm{X}}_V + \E{\bm{X}|\bm{Y}}$.

%%%%%%%%%%%%%%%%%%%%%%%%%
\subsection{Wyner-Ziv Rates}
\label{subsec:WZ-rates}
The WZ extension of RD theory assumes the decoder, but not the encoder, has access to $\bm{Y}$. The RD function is now
\begin{align}
    R_{\text{WZ}}(\mathcal D) = \min_{\E{d(\bm{X},\hat{\bm{X}})}\le \mathcal D} \left( I(\bm{U};\bm{X}) - I(\bm{U};\bm{Y}) \right )
    \label{eq:WZ-RD-function}
\end{align}
where the minimization is over all $\bm{U}$ such that $\bm{U}\leftrightarrow \bm{X}\leftrightarrow \bm{Y}$ forms a Markov chain, and all functions $f(\cdot)$ such that $\hat{\bm{X}} = f(\bm{U},\bm{Y})$. The Markov chain relation gives
\begin{align}
    I(\bm{U};\bm{X}) - I(\bm{U};\bm{Y})
    = I(\bm{U} ; \bm{X} | \bm{Y}).
    \label{eq:R-diff-cond}
\end{align}

Remarkably, $R_{\text{WZ}}(\mathcal D)=R_{\bm{X}|\bm{Y}}(\mathcal D)$ if $\bm{X},\bm{Y}$ are jointly Gaussian and $d(\bm{x},\hat{\bm{x}})=\|\bm{x}-\hat{\bm{x}}\|^2$, i.e., the RD function is the same as if the encoder also has $\bm{Y}$. For example, consider the scalar problem and choose $U=X+\check{Z}$ where $\check{Z}$ is independent of $(X,Y)$. The best estimate of $X$ is
\begin{align}
    \hat{X} = \E{X|U,Y}
    = \frac{\sigma_{x|y}^2\, U + \sigma_{\check{z}}^2\, \E{X|Y}}{\sigma_{x|y}^2+\sigma_{\check{z}}^2}.
    \label{eq:WZ-estimate}
\end{align}
If $0<\mathcal{D}< \sigma_{x|y}^2$, choose the description noise variance as
\begin{align}
    \sigma_{\check{z}}^2
    = \frac{\sigma_{x|y}^2 \mathcal D}{\sigma_{x|y}^2 - \mathcal D} 
    \implies
    \mathcal{D} = \frac{\sigma_{x|y}^2 \sigma_{\check{z}}^2}{\sigma_{x|y}^2 + \sigma_{\check{z}}^2}
    \label{eq:WZ-variances}
\end{align}
This choice gives $\E{(X-\hat{X})^2}=\mathcal{D}$ and
\begin{align}
    I(U;X) - I(U;Y)
    %& = \frac{1}{2} \log \frac{\sigma_x^2+\sigma_{\check{z}}^2}{\sigma_{\check{z}}^2} - \frac{1}{2}\log \frac{\sigma_x^2+\sigma_{\check{z}}^2}{\sigma_{x|y}^2+\sigma_{\check{z}}^2} \nonumber \\
    %& = h(U|Y) - h(U|X) \nonumber \\
    & = \frac{1}{2}\log(\sigma_{x|y}^2\, \big/ \mathcal D).
    \label{eq:WZ-rate-scalar}
\end{align}
 
%%%%%%%%%%%%%%%%%%%%%%%%%
\section{Scalar Models and Quantization}
\label{sec:scalar-models}

\begin{figure}[!t]
\centering
%auto-ignore
\begin{tikzpicture}[scale=0.75, every node/.style={scale=0.75}]
\coordinate (a) at (2.5,3.2);
\node[above = 1pt of a] {$\bm{x}$};

\coordinate (b) at (3.3,0.2);
\node[above = 0.5pt of b] {$\bm{d}$};

\coordinate (c) at (5.1,2);
\node[above = 1pt of c] {$w$};

%dither lines
\draw [thick](1,0.2) -- (7.8,0.2);
\draw [-latex,thick](3,0.2) -- (2,0.2);
\draw [-latex,thick](3,0.2) -- (4.2,0.2);
\draw [-latex,thick](1,0.2) -- (1,1.86);
\draw [-latex,thick](6.2,0.2) -- (6.2,1.68);
\draw [-latex,thick](3,0.7) -- (3,0.2);

% X-Y channel
\node[thick] (ee) at (2.99,3.2){$\bigoplus$};

% p(Y|X) channel lines
\draw [thick](1,3.2)--(1,2.81);
\draw [-latex,thick](1,2.5)--(1,2.15);
\draw [-latex,thick](2.82,3.2)--(2,3.2);
\draw [thick](2.1,3.2)--(1,3.2);

\draw [-latex,thick](6.2,3.2)--(3.16,3.2);
\draw [-latex,thick](3.16,3.2)--(5,3.2);

\draw [-latex,thick](6.2,3.2) -- (6.2,2.3);

\draw [-latex,thick](3,3.9) -- (3,3.38);
\draw [-latex,thick](4.1,3.7) -- (4.1,3.2);

%modulo and encoder input lines
\draw [-latex,thick](1.2,2) -- (1.56,2);
\draw [-latex,thick](2.8,2) -- (3.4,2);
\node[thick] (aaa) at (1,2){$\bigoplus$};

%m decoder line
\draw [-latex,thick](4.83,2) -- (5.48,2);
\coordinate (c2) at (4.6,1.76);

\node[thick] (aa) at (7.8,2){$\bigoplus$};

% lines to plus symbol
\draw [thick](6.2,3.2) -- (8,3.2);
\draw [-latex,thick](7.8,3.2) -- (7.8,2.81);
\draw [-latex,thick](7.8,2.5) -- (7.8,2.15);
\draw [-latex,thick](7.8,0.2) -- (7.8,1.86);
\draw [-latex,thick](6.91,2) -- (7.62,2);

%line to mod a
%\draw [-latex,thick](8.16,1.981) -- (7,1.981);

\node[thick] (bb) at (10.4,2){$\bigotimes$};
\node[thick] (cc) at (7.8,2.65){$\bigotimes$};
\node[thick] (ccc) at (1,2.65){$\bigotimes$};

%line to mod A
\draw [-latex,thick](7.99,2) -- (8.35,2);

%line to mult
\draw [-latex,thick](9.6,2) -- (10.22,2);
\draw [-latex,thick](10.58,2.01) -- (10.96,2.01);

\node[thick] (dd) at (11.1,2){$\bigoplus$};

\draw [-latex,thick](8,3.2) -- (9.1,3.2);
\draw [thick](9,3.2) -- (11.1,3.2);
\draw [-latex,thick](11.1,3.2) -- (11.1,2.16);

% line to x_hat
\draw [-latex,thick](11.28,2) -- (11.68,2);

\coordinate (d) at (4.1,2.6);
\node[above = 1pt of d] {$\bm{y}$};
\coordinate (e) at (7.15,2);
\node[above = 1pt of e] {$\bm{u}$};
\coordinate (f) at (8.13,2.4);
\node[above = 2pt of f] {$\alpha$};

\coordinate (g) at (8,2);
\node[above = 1pt of g] {-};

\coordinate (h) at (8,1.5);
\node[above = 1pt of h] {-};

\coordinate (i) at (10.4,2.1);
\node[above = 1pt of i] {$\alpha$};
\node[below = 4pt of i, xshift=-0.5cm] {$\bm{z}'$};

\coordinate (ii) at (1,2.7);
\node[right = 2pt of ii] {$\alpha$};
\coordinate (ij) at (3.1,2);
\node[above = 1pt of ij] {$\bm{x}'$};

\coordinate (j) at (11.8,1.76);
\node[above = 1pt of j] {$\hat{\bm{x}}$};

\coordinate (k) at (3,3.83);
\node[above = 1pt of k] {$\bm{z}$};

\node [draw, shape=rectangle,thick, minimum width= 1.0cm, minimum height=0.6cm, text width=0.6cm, anchor=center] at (4.1,4) {$p(y)$};
\node [draw, shape=rectangle,thick, minimum width= 1.0cm, minimum height=0.6cm, text width=0.6cm, anchor=center] at (3,1.02) {$p(d)$};
\node [draw, shape=rectangle,thick, minimum width= 1cm, minimum height=0.6cm, text width=1cm, anchor=center] at (2.16,2) {mod~$A$};
\node [draw, shape=rectangle,thick, minimum width= 1cm, minimum height=0.6cm, text width=1cm, anchor=center] at (8.98,2) {mod~$A$};
\node [draw, shape=rectangle,thick, minimum width= 1.2cm, minimum height=0.6cm, text width=1.2cm, anchor=center] at (4.1,2) {Encoder};
\node [draw, shape=rectangle,thick, minimum width= 1.2cm, minimum height=0.6cm, text width=1.2cm, anchor=center] at (6.2,2) {Decoder};

\end{tikzpicture}
\caption{WZ coding with a modulo operator and dithering.}
\label{fig:DPC-lattice}
\end{figure}
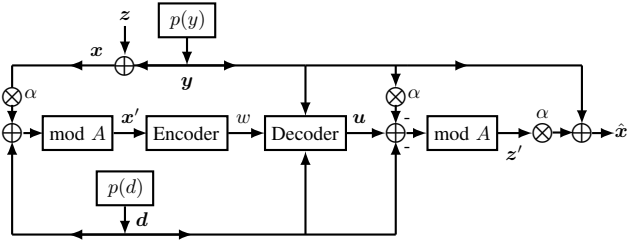

Consider $X=Y+Z$ where $Z$ is Gaussian and independent of $Y$, i.e., we have $\sigma_{x|y}^2=\sigma_z^2$. Fig.~\ref{fig:DPC-lattice} shows the WZ coding structure of \cite[Sec.~IV.A]{zamir2002nested} but with scalar operations. The encoder and decoder share a dither $D$ statistically independent of $(X,Z)$ and uniformly distributed in $[-A/2, A/2)$. The inflation factor is $\alpha=\sqrt{1-\sigma_d^2/\sigma_z^2}$, where $\sigma_d^2$ is a target distortion; cf. \cite[p.~1260]{zamir2002nested}.

%%%%%%%%%%%%%%%%%%%%%%%%
\subsection{Model in a Finite Interval}
\label{subsec:derived-model}
We derive a model for the interval $[-A/2,A/2)$ by modifying steps in \cite{zamir2002nested,Sener-CL21,Sener-CL24,Sener-ISIT24}. The derived source is
\begin{align}
    (X',Y')^T = \left( \alpha\, (X,Y)^T + D\cdot\bm{1}_2 \right)\ \textrm{mod}\ A\cdot\bm{1}_2 .
    \label{eq:derived-source}
\end{align}
The dither makes $X'$ uniform over $[-A/2, A/2)$ and independent of $X$, and similarly for $Y'$ and $Y$. The source $X'$ is quantized via $U$ with discrete alphabet $\mathcal{U}$. Using standard theory, the decoder can recover $U$ if the rate satisfies
\begin{align}
    R = I(U;X') - I(U;Y')
    \label{eq:WZ-rate}
\end{align}
where $U\leftrightarrow X'\leftrightarrow Y'$ forms a Markov chain; see \eqref{eq:WZ-RD-function}.

Let $\sigma_d^2$ be a distortion parameter satisfying $0 < \sigma_d^2 < \sigma_z^2$.
Define the encoder and decoder noise
\begin{align}
    \tilde{Z} & = (U - X')\ \textrm{mod}\ A
    \label{eq:Ztilde} \\
    Z' & = (U - Y')\ \textrm{mod}\ A
    = \big( \tilde Z  + \alpha Z \big)\ \textrm{mod}\ A
   \label{eq:Zprime}
\end{align}
where $U$ is selected based on $X'$ only; see Sec.~\ref{subsec:lattices-and-shaping} below. Thus, $U\leftrightarrow X'\leftrightarrow Y'$ forms a Markov chain, as required.

The decoder puts out the reconstruction
\begin{align}
    \hat{X} = Y + \alpha Z'
    \label{eq:reconstruction}
\end{align}
Observe that $\hat{X} = \E{X|Y,Z'}$ if $Z'=\tilde Z + \alpha Z$ is Gaussian and $\tilde Z$ is independent of $(Y,Z)$ and has variance $\sigma_d^2$. Also, $\hat X$ is a function of $U$ and $Y$, as required.

%-------------------------------
\subsection{Scalar Lattices and Probabilistic Shaping} 
\label{subsec:lattices-and-shaping}

Consider the equally spaced scalar quantizer with alphabet
\begin{equation}
 \mathcal{U} = \big\{-A/2 + (2k+1)\kappa\ \big\}_{k=0}^{M-1}
 \label{eq:constellation}
\end{equation}
where $A\ge0$ and $\kappa = A/(2M)$.
We call this alphabet $M$-ASK to match \cite{Sener-CL21,Sener-CL24,Sener-ISIT24}.
%Observe that $\mathcal{U}$ is a subset of the scalar lattice $\mathbb{Z} \cdot A/M$ if $M$ is odd and $(2\mathbb{Z}+1) \cdot A/(2M)$ if $M$ is even.
%
Let $q(\tilde z)$ be a density on $\tilde{z} \in [-A/2,A/2)$ and apply the probabilistic shaping
\begin{equation}
    P(u|x') = \frac{q\big((u-x')\ \text{mod}\ A\big)}{\sum_{v\in\mathcal U} q\big((v-x')\ \text{mod}\ A\big)}, \quad u \in \mathcal U .
\label{eq:q_shaping}
\end{equation}
Observe that $X'$, $\tilde Z$ play the roles of $S'$, $X$ in \cite{Sener-CL24}, e.g.,
the continuously uniform $X'$ induces a discrete uniform $U$. To see this, define $\tilde{x}=(u-x')\ \textrm{mod}\ A$ and
\begin{align}
    d(x) = 2 \kappa \sum\nolimits_{k=0}^{M-1} q((x + 2 k \kappa)\ \text{mod}\ A).
\end{align}
We then have
\begin{align}
    P(u)
    = \int_{-A/2}^{A/2} \frac{1}{A}\, P(u|x')\, dx'
    = \int_{-A/2}^{A/2} \frac{1}{M}\, \frac{q(\tilde x)}{d(\tilde x)} \, d\tilde x
    \label{eq:Pu}
\end{align}
and $P(u)=1/M$ for all $u$ since \eqref{eq:Pu} does not depend on $u$.

We claim $U,\tilde{Z},Z$ are mutually statistically independent and
\begin{align}
    p(\tilde z) = q(\tilde z)/d(\tilde z),
    \;\; \tilde z \in [-A/2,A/2).
    \label{eq:pz}
\end{align}
To see this, let $x'=(u-\tilde z)\ \textrm{mod}\ A$. We have
\begin{align}
    p(\tilde z,z|u)
    & = p(x',z|u) \nonumber \\
    %& \overset{(a)}{=} p(x'|u)\, p(z) \nonumber \\
    & \overset{(a)}{=} \left( p(x') P(u|x') / P(u) \right) \, p(z) \nonumber \\
    & \overset{(b)}{=} \frac{q(\tilde z)}{d(\tilde z)}\, p(z)
\end{align}
where step $(a)$ follows because $(X',U)$ and $Z$ are independent, and step $(b)$ follows by $p(x')=1/A$ and $P(u)=1/M$.

%-------------------------------
\subsection{Truncated Gaussian Shaping} 
We use truncated Gaussian shaping with (see~\cite[eq.~(12)]{Sener-CL24})
\begin{equation}
q(\tilde{z}) = \frac{e^{-\tilde{z}^2 / (2\sigma_d^2)}}{c \cdot \sqrt{2\pi\sigma_d^2}}, \quad \tilde{z} \in [-A/2,A/2)
\label{eq:TG-shaping}
\end{equation}
where $c=1-2Q(A/(2\sigma_d))$. We compute (see~\cite[eq.~(16)]{Sener-CL24})
\begin{align}
    P_{q,\tilde Z} := \Eq{\tilde{Z}^2}
    = \sigma_d^2 \left[ 1 - \frac{A e^{-A^2/(8\sigma_d^2)}}{c \sqrt{2 \pi \sigma_d^2}} \right]
    \label{eq:Ztilde-variance}
\end{align}
and (see~\cite[eq.~(17)]{Sener-CL24})
\begin{align}
    h_q(\tilde Z )
    = \frac{1}{2} \log\left(2 \pi e \sigma_d^2 c^2 \right) - \frac{1}{2} \left( 1 - \frac{P_{q,\tilde Z}}{\sigma_d^2} \right).
    \label{eq:Ztilde-entropy}
\end{align}
Since $q(.)$ is symmetric unimodal, by
\cite[eq.~(20)]{Sener-ISIT24} we have
$d_\textrm{min} \le d(x) \le d_\textrm{max}$ where 
\begin{align}
    d_\textrm{min} = 1 - \frac{A}{M} q(0), \quad 
    d_\textrm{max} = 1 + \frac{A}{M} q(0).
\label{eq:D-extrema}
\end{align}

%-------------------------------
\subsection{Distortion} 
\label{subsec:distortion-scalar}
To bound the distortion, consider
 \begin{align}
    X - \hat{X}
    & = Z - \alpha Z'. %\nonumber \\
    %& = Z - \alpha \left( (\tilde Z  + \alpha Z)\ \textrm{mod}\ A \right)
\end{align}
The modulo operator cannot increase the second moment, so
\begin{align}
    \E{(Z')^2}
    & \le \E{\big({\tilde Z} + \alpha Z\big)^2}  \nonumber \\
    & \overset{(a)}{\le} \frac{1}{d_\textrm{min}} \Eq{{\tilde Z}^2}
    + \alpha^2 \sigma_z^2 \nonumber \\
    & = \sigma_z^2 + \big(P_{q,\tilde Z} \big/ d_\textrm{min} - \sigma_d^2 \big)
    \label{eq:Zprime-variance-bound2}
\end{align}
where step $(a)$ follows by $p(\tilde z)\le q(\tilde z)/d_\textrm{min}$ if $M$ is sufficiently large so $d_\textrm{min}>0$; see \eqref{eq:pz} and \eqref{eq:D-extrema}. We further have
\begin{align}
    \E{ Z Z'}
    & = \E{ Z \big( \tilde Z + \alpha Z - I A \big)} \nonumber \\
    & = \alpha\, \sigma_z^2 - A\, \E{I Z}
    \label{eq:cross-correlation}
\end{align}
where we used the independence of $Z$ and $\tilde Z$, and $I$ is the RV taking on the modulo shift values $i$. We thus have
\begin{align}
    & \E{\big(X - \hat X\big)^2}
    = \sigma_z^2
    - 2 \alpha\, \E{Z Z'}
    + \alpha^2\,\E{(Z')^2} \nonumber \\
    & \overset{(a)}{\le}
    %\left(1-\alpha^2\right) \sigma_z^2
    %+ 2 \alpha A \, \E{iZ} 
    %+ \alpha^2 \left(P_{q,\tilde Z} \big/ d_\textrm{min} - %\sigma_d^2 \right)
    %\nonumber \\ & =
    \sigma_d^2
    + 2 \alpha A \, \E{I Z}
    + \alpha^2 \big(P_{q,\tilde Z} \big/ d_\textrm{min} - \sigma_d^2 \big)
    \label{eq:distortion}
\end{align}
where step $(a)$ follows by \eqref{eq:Zprime-variance-bound2} and \eqref{eq:cross-correlation}.

We bound the term $2 \alpha A\,\E{I Z}$. Define the intervals
\begin{align}
   \mathcal I_k = \left[ kA - A/2, kA + A/2 \right), \;\; k\in\mathbb Z
   \label{eq:intervals}
\end{align}
and let $Z''=\tilde Z + \alpha Z$. By symmetry, we have
\begin{align}
    \E{Z \big| Z''\in\mathcal I_{(-k)} }
    = -\E{Z \big| Z''\in\mathcal I_k }.
\end{align}
The law of total expectation thus gives
\begin{align}
    2 \alpha A\,\E{I Z} & = 4 \alpha \sum\nolimits_{k\ge 1} \Prob{Z''\in\mathcal I_k}\, k A\,
    \underbrace{\E{Z \big| Z''\in\mathcal I_k }}_{\displaystyle \le kA/\alpha} \nonumber \\
    %& \le 4 \sum\nolimits_{k\ge 1} \Prob{Z''\in\mathcal I_k}\, (kA)^2 \nonumber \\
    & \le 4 \int_{A/2}^\infty p_{Z''}(x) \cdot (x+A/2)^2\, dx
    \label{eq:EiZ-bound}
\end{align}
where
\begin{align}
    p_{Z''}(x) & = p_{\tilde Z}*p_{\alpha Z}(x) %\nonumber \\
    %& 
    = \int_{-A/2}^{A/2} \frac{q(y)}{d(y)}\, p_{\alpha Z}(x-y)\, dy \nonumber \\
    & \overset{(a)}{\le} \frac{e^{-x^2/(2\sigma_z^2)}}{c\,d_{\textrm{min}}\cdot \sqrt{2 \pi \sigma_z^2}}
\end{align}
and where step $(a)$ follows by extending the domain of $q(\cdot)$ in \eqref{eq:TG-shaping} to all reals. We may thus bound \eqref{eq:EiZ-bound} as
\begin{align}
    & 2 \alpha A\, \E{I Z}
    \le \frac{4}{c\,d_{\textrm{min}}} \int_{A/2}^\infty \frac{e^{-x^2/(2\sigma_z^2)}}{\sqrt{2 \pi \sigma_z^2}} \cdot (x+A/2)^2\, dx \nonumber \\
    & = \frac{4}{c\,d_{\textrm{min}}} \left[  \frac{3A\,\sigma_z^2}{2} \cdot \frac{e^{-A^2/(8\sigma_z^2)}}{\sqrt{2\pi\sigma_z^2}} + \left(\sigma_z^2 + \frac{A^2}{4}\right) Q\left(\frac{A}{2\sigma_z}\right) \right] .
    \label{eq:EiZ-bound2}
\end{align}
\begin{comment}
where we used
\begin{align}
    %\int_x^\infty y\, e^{-y^2/2}\, dy & = e^{-x^2/2} \\
    \int_x^\infty y^2\, e^{-y^2/2}\, dy & = x\, e^{-x^2/2} + \sqrt{2\pi}\, Q(x).
\end{align}
\end{comment}
The right-hand side of \eqref{eq:EiZ-bound2} vanishes as $A\rightarrow\infty$.

%-------------------------------
\subsection{Source Coding Rate} 
For the rate $R$, similar to \cite{Sener-CL24}, we have
\begin{align}
   I(U;X') & = h(X') - h(\tilde Z|U) = \log A - h(\tilde Z)
   \label{eq:message-and-shaping-rate}
   \\
   I(U;Y') & = h(Y') - h(Z'|U) = \log A - h(Z')
   \label{eq:shaping-rate}
\end{align}
where we used the independence of $U$, $\tilde Z$, $Z$. We may use \cite[eq.~(25)]{Sener-ISIT24} to lower bound $h(\tilde Z)$ and \eqref{eq:Zprime-variance-bound2} gives
\begin{align}
    h(Z')\le \frac{1}{2}\log\left(2\pi e \big(\sigma_z^2 + P_{q,\tilde Z} \big/ d_\textrm{min}-\sigma_d^2\big)\right) .
    \label{eq:hZprime-bound}
\end{align}
We proceed as in \cite{Sener-CL24,Sener-ISIT24} and first take the limit $M\rightarrow\infty$ to obtain $d(\tilde z)\rightarrow 1$ and $p(\tilde z)\rightarrow q(\tilde z)$ for $\tilde z \in [-A/2,A/2)$. We next take the limit $A\rightarrow \infty$ to obtain $p(\tilde z) \rightarrow \mathcal N(0;\sigma_d^2)$ and $p(z') \rightarrow \mathcal N(0;\sigma_z^2)$. We therefore
\begin{comment}
have
\begin{align}
    & \lim_{A\rightarrow\infty} \lim_{M\rightarrow\infty} h(\tilde Z) \rightarrow \frac{1}{2} \log(2 \pi e \sigma_d^2) \\
    & \lim_{A\rightarrow\infty} \lim_{M\rightarrow\infty} h(Z') \rightarrow \frac{1}{2} \log(2 \pi e \sigma_z^2)
\end{align}
and
\end{comment}
obtain the WZ rate.

%-------------------------------
\subsection{Simulations with Short Polar Codes}
\label{subsec:scalar-simulations}
\begin{figure}[!t]
\centering
\input{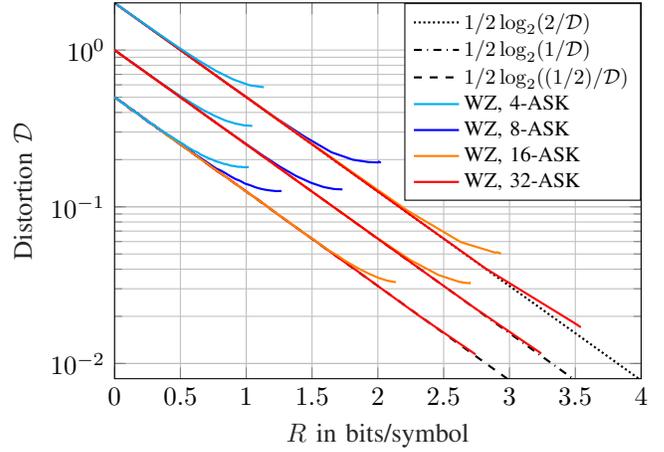}
\caption{RD curves for $(\sigma_y^2,\sigma_z^2)=(0,2)$, $(1,1)$, $(3/2,1/2)$ and $\sigma_x^2=2$. The target rates are $R=\frac{1}{2}\log(\sigma_z^2/\mathcal D)$.
}
\label{fig:rates}
\end{figure}

Fig.~\ref{fig:rates} shows RD curves for $M=4,8,16,32$. The 4-ASK curves have $A=10,8,6$ for $\sigma_y^2=0,1,3/2$, respectively. The 8-ASK and 16-ASK curves have $A=12,10,10$, and the 32-ASK curves have $A=14,12,10$. The figure illustrates that increasing $M$ and $A$ achieves the WZ curve.

Consider $R=1$ and 8-ASK with natural labeling. We use multilevel coding with $L=\log_2 M$ levels, where levels 1 to $L$ carry successively more significant bits. Each level has one 5G NR polar code of length 256. The encoder and decoder use successive cancellation list decoding \cite{Tal-Vardy-IT15} with list size 8 and list passing across levels \cite{Prinz-Yuan-ISTC18,Karakchieva-SCC19}. The most reliable polar subchannels of each level are assigned shaping bits, the next most reliable information bits, and the rest frozen bits. We optimized $A$, $\sigma_d^2$, and the numbers of message and shaping bits across the polar subchannels to minimize $\E{\mathcal D}$.
%For example, PCQmod with 8-ASK has the following (message,shaping,frozen) bit numbers: level 1 has (19,0,237); level 2 has (176,19,61); level 3 has (61,195,0).
%

\begin{figure}[!t]
\centering
\input{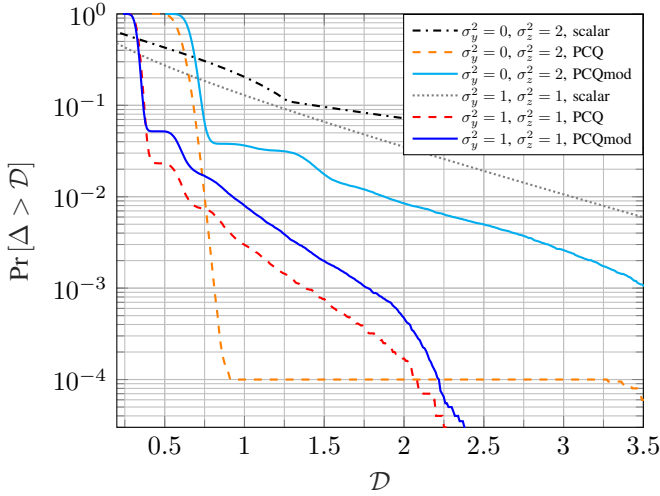}
\caption{$\Prob{\Delta>\mathcal D}$ for $\sigma_x^2=2$, $R=1$, 8-ASK, and $n=256$.}
\label{fig:CDFplot}
\end{figure}

We study two sources; Fig.~\ref{fig:CDFplot} plots $\Prob{\Delta>\mathcal D}$.
\vspace{-0.1cm}
\begin{itemize}[leftmargin=*]
\item Source 1: $\sigma_y^2=0$, $\sigma_z^2=\sigma_x^2=2$, $\E{\Delta}\ge 0.5$
    \begin{itemize} \itemsep 0pt
    \item one-bit scalar quantizer: $\E{\Delta}=0.727$
    \item PCQ: $\E{\Delta}=0.614$
    for ASK spacing $0.6$
    %; 16-ASK, $\E{\Delta}=0.733$
    \item PCQmod: $\E{\Delta}=0.713$ for $A/M=1.5$
    %\item \textcolor{red}{PCQmod and 16-ASK: $\E{\Delta}=0.???$ for $A/M=?.??$}
    \end{itemize}
\item Source 2: $\sigma_y^2=\sigma_z^2=1$, $\E{\Delta}\ge 0.25$
    \begin{itemize} \itemsep 0pt
    \item one-bit scalar quantizer: $\E{\Delta}=0.519$
    \item PCQ: $\E{\Delta}=0.351$
    for ASK spacing $0.8$
    %; 16-ASK, $\E{\Delta}=0.358$
    \item PCQmod: $\E{\Delta}=0.357$ for $A/M=1.25$
    \end{itemize}
\end{itemize}
\vspace{0.1cm}
Fig.~\ref{fig:CDFplot} shows PCQ outperforms PCQmod, notably for $\sigma_y^2=0$ where PCQ can use a (relatively) fine ASK spacing with few shaping bits. PCQmod requires coarser ASK spacing and more shaping bits because $A$ must be sufficiently large to limit the number of modulo shifts, while the dither makes $I(U;Y')$ grow as $\log A$ even if $Y=0$; see~\eqref{eq:shaping-rate}. Fig.~\ref{fig:CDFplot} also shows PCQ and PCQmod perform similarly as $\sigma_y^2$ increases; this is because WZ coding requires shaping bits for both methods, as represented by $I(U;Y)$ for PCQ and $I(U;Y')$ for PCQmod. We remark dithering benefits quantization and secrecy by making $(X',M)$ independent of $X$ \cite[Ch.~4]{zamir14}.

%%%%%%%%%%%%%%%%%%%%%%%%%
\section{Vector Models and Quantization}
\label{sec:vector-models}
Consider the vectors $\bm{X},\bm{Y}$. We re-define (see \eqref{eq:XV})
\begin{align}
    \bm{X}_V := V^T \bm{X} = V^T \E{\bm{X}|\bm{Y}} + V^T \bm{Z}
    \label{eq:vector-source-transformed}
\end{align}
because the encoder does not have $\bm{Y}$. Let $\bm{Y}_V = V^T \E{\bm{X}|\bm{Y}}$ so we have (see \eqref{eq:QXV})
\begin{align}
    Q_{\bm{X}_V|\bm{Y}}
    = Q_{\bm{X}_V|\bm{Y}_V}
    = Q_{\bm{V}^T \bm{Z}} = \Lambda .
\end{align}
Thus, the $X_{V,1},\dots,X_{V,\ddim}$ are independent given $\bm{Y}$ or $\bm{Y}_V$ . Moreover, if $\bm{v}_i^T$ is the $i$-th row of $V^T$, we have
\begin{align}
    \E{X_{V,i}|\bm{Y}}
    & = \bm{v}_i^T \E{\bm{X} \bm{Y}^T} Q_{\bm{Y}}^{-1} \bm{Y}
    = Y_{V,i} %\nonumber \\
    %\E{X_{V,i}|Y_{V,i}}
    %& = \frac{\E{X_{V,i} Y_{V,i}}}{\E{Y_{V,i}^2}} Y_{V,i}
    %= Y_{V,i} 
\end{align}
and therefore $X_{V,i}\leftrightarrow Y_{V,i}\leftrightarrow \bm{Y}$ forms a Markov chain.

For coding, mimic \eqref{eq:WZ-estimate}--\eqref{eq:WZ-rate-scalar} and choose
\begin{align}
    U_i=X_{V,i}+\check{Z}_i, \quad i=1,\dots,\ddim
\end{align}
where $\check{Z}_1,\dots,\check{Z}_n$, $(\bm{X},\bm{Y})$ are mutually independent. We have the Markov chains $U_i\leftrightarrow
X_{V,i}\leftrightarrow Y_{V,i} \leftrightarrow\bm{Y}$ and \eqref{eq:R-diff-cond} gives
\begin{align}
    R = I(\bm{U};\bm{X}|\bm{Y})
    %& = I(\bm{U};\bm{X}_V|\bm{Y}_V) \nonumber \\
    & = \sum\nolimits_{i=1}^\ddim I(U_i;X_{V,i}|Y_{V,i})
    \label{eq:WZ-sum}
\end{align}
where $X_{V,i} -\E{X_{V,i}|Y_{V,i}}$ has variance $\lambda_i$. We describe $X_{V,i}$ with the reverse waterfilling level $\lambda$ as in Sec.~\ref{subsec:WZ-rates} to obtain the $\hat{X}_{V,i}$ in \eqref{eq:forward-channel-i}. Thus, the distortions for $i=1,\dots,\ddim$ are $\min(\lambda_i,\lambda)$ and the sub-channel rates are
\begin{align}
    R_i = I(U_i;X_{V,i}|Y_{V,i}) = \left\{
    \begin{array}{ll}
    \frac{1}{2}\log(\lambda_i/\lambda), & \lambda < \lambda_i \\
    0, & \lambda \ge \lambda_i
    \end{array} \right. .
\end{align}

%-------------------------------
\subsection{Simulations with Short Polar Codes} 
\label{subsec:vector-simulations}

\begin{figure}[!t]
\centering
\input{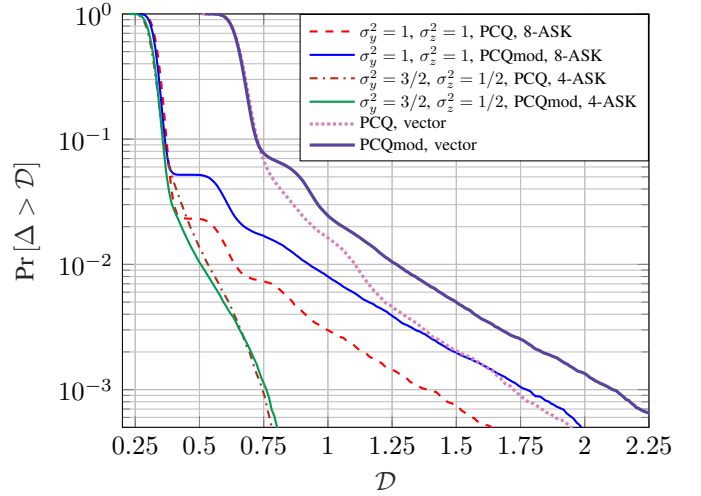}
\caption{$\Prob{\Delta>\mathcal D}$ for a vector source, $R=3/2$, and $n=256$.
\vspace{-0.2cm}
}
\label{fig:CDFplot2}
\end{figure}

The expression \eqref{eq:WZ-sum} suggests using the scalar structure of Sec.~\ref{sec:scalar-models} in parallel with different $A_i$, $M_i$ for each source pair $X_{V,i},Y_{V,i}$, $i=1,\dots,\ddim$. 
For example, consider $\ddim=2$ and 
\begin{align}
    Q_{\bm{Z}} = \frac{1}{4} \begin{pmatrix}
        3 & 1 \\ 1 & 3
    \end{pmatrix}  = V^T\Lambda V
    \label{eq:vector-example-QZ}
\end{align}
where
\begin{align}
    \Lambda = \begin{pmatrix}
        1 & 0 \\ 0 & 1/2
    \end{pmatrix},\quad
    V = \frac{1}{\sqrt{2}} \begin{pmatrix}
        1 & 1 \\ 1 & -1
    \end{pmatrix} .
\end{align}
Suppose $\E{Y_{V,1}^2}=1$ and $\E{Y_{V,2}^2}=3/2$. 
The rate $R=3/2$ is achieved with $\lambda=1/4$ and $(R_1,R_2)=(1,1/2)$, i.e., both sub-channel encoders are active with target distortion $\lambda=1/4$ and $\E{\Delta}\ge2\lambda=1/2$. 
We study 8-ASK for $R_1=1$ and 4-ASK for $R_2=1/2$. Let $\Delta_1$ and $\Delta_2$ be the corresponding distortion RVs. We have $R=R_1+R_2$ and $\Delta=\Delta_1+\Delta_2$. Fig.~\ref{fig:CDFplot2} plots the $\Prob{\Delta>\mathcal D}$ curves; cf. Fig.~\ref{fig:CDFplot}.

%\vspace{-0.1cm}
\begin{itemize}[leftmargin=*]
\item Source 1: $\sigma_y^2=\sigma_z^2=1$, $R_1=1$, $\E{\Delta_1}\ge 0.25$
    \begin{itemize} \itemsep 0pt
    \item PCQ: $\E{\Delta_1}=0.351$ for ASK spacing $0.8$
    \item PCQmod: $\E{\Delta_1}=0.357$ for $A/M=1.25$
    \end{itemize}
\item Source 2: $\sigma_y^2=3/2$, $\sigma_z^2=1/2$, $R_2=1/2$, $\E{\Delta_2}\ge 0.25$
    \begin{itemize} \itemsep 0pt
    \item PCQ: $\E{\Delta_2}=0.327$ for ASK spacing $1.5$
    \item PCQmod: $\E{\Delta_2}=0.328$ for $A/M=1.5$
    \end{itemize}
%\item Vector Source: $R=3/2$, $\E{\Delta}\ge 0.5$
%    \begin{itemize} \itemsep 0pt
%    \item PCQ: \textcolor{red}{$\E{\Delta}=0.???$}
%    \item PCQmod: \textcolor{red}{$\E{\Delta}=0.???$}
%    \end{itemize}
\end{itemize}
The vector source has $R=3/2$ and $\E{\Delta}=\E{\Delta_1}+\E{\Delta_2}$.

%For the $R = 1/2$ code, (message,shaping,frozen) bit numbers are as follows. Level 1 has (76,20,160); level 2 has (52,199,5). For the $R = 1$ code, level 1 has (49,0,207); level 2 has (181,50,25); level 3 has (26,230,0).

\begin{comment}
Observe that we have
\begin{align}
    Q_{\bm{Y}_V}
    & = V^T\left( Q_{\bm X} - Q_{\bm Z} \right) V \nonumber \\
    & = V^T \E{\bm{X} \bm{Y}^T} Q_{\bm{Y}}^{-1} \E{\bm{Y} \bm{X}^T} V.
\end{align}
For example, suppose $\bm{X}=[a\;b]^T Y + \bm{Z}$. We then have
\begin{align}
    Q_{\bm{Y}_V}
    & = \frac{\sigma_y^2}{2} \begin{pmatrix} a+b \\ a-b \end{pmatrix}
    \begin{pmatrix} a+b & a-b \end{pmatrix} . %\nonumber \\
    %& = \frac{\sigma_y^2}{2} \begin{pmatrix}
    %    (a+b)^2 & a^2-b^2 \\ a^2-b^2 & (a-b)^2
    %\end{pmatrix} .
    \label{eq:vector-example-QYV}
\end{align}
Now suppose $\E{Y_{V,1}^2}=1$, $\E{Y_{V,2}^2}=3/2$. 
For example, choose $\sigma_y^2=1$ and
\begin{align}
    a = \big(\sqrt{2}+\sqrt{3}\,\big) \big/2, \quad
    b = \big(\sqrt{2}-\sqrt{3}\,\big) \big/2
\end{align}
in \eqref{eq:vector-example-QYV} to obtain
\begin{align}
    Q_{\bm{Y}_V}
    & = \begin{pmatrix}
        1 & \sqrt{3/2}\text{\:} \\ \sqrt{3/2} & 3/2
    \end{pmatrix} .
    \label{eq:vector-example-QYV2}
\end{align}
\end{comment}

%%%%%%%%%%%%%%%%%%%%%%%%%
\section{Conclusions}
\label{sec:conclusions}
We showed PCQmod achieves the WZ curve for Gaussian vector sources and mean square error distortion. Vector coding uses a parallel structure motivated by \eqref{eq:WZ-sum} with the transformed source $\bm{X}_V,\bm{Y}_V$. Simulations with short polar codes show that PCQ outperforms PCQmod without side information. PCQ and PCQmod perform similarly as the side information becomes prominent, and PCQmod has the benefits of dithering.

Future work could replace the decoder's modulo operator with a scalar minimum mean square error estimator. One could also study different dynamic ranges of dithering, both with and without modulo operators, and how dithering provides robustness for realistic sources, i.e., non-Gaussian $\bm{X},\bm{Y}$.

%\clearpage
%%%%%%%%%%%%%%%%%%%%%%
\section{Acknowledgements}

This work was supported by the German Research Foundation (DFG) via the German-Israeli Project Cooperation (DIP) under projects KR 3517/13-1 and SH 1937/1-1.

\clearpage
%%%%%%%%%%%%%%%%%%%%%%%%%%%%%
\bibliographystyle{IEEEtran}
\bibliography{references}

\end{document}